# Development of a cosmic ray oriented trigger for the fluorescence telescope on EUSO-SPB2

George Filippatos [a,*], Matteo Battisti [b,c,*], Alexander Belov [d,e], Mario Bertaina [b,c], Francesca Bisconti [b,c], Johannes Eser [f], Marco Mignone [c], Fred Sarazin [a], Lawrence Wiencke [a]

[a] *Colorado School of Mines, 1500 Illinois St, Golden, CO 80401, USA*
[b] *Universitá degli Studi di Torino via P. Giuria 1, 10125 Torino, Italy*
[c] *Istituto Nazionale di Fisica Nucleare, sez. di Torino, via P. Giuria 1, 10125 Torino, Italy*
[d] *M.V. Lomonosov Moscow State University, Faculty of Physics, Ulitsa Kolmogorova, 1c2, Moscow 119234, Russia*
[e] *M.V. Lomonosov Moscow State University, D.V. Skobeltsyn Institute of Nuclear Physics, Ulitsa Kolmogorova, 1c2, Moscow 119991, Russia*
[f] *University of Chicago, 5640 South Ellis Avenue, Chicago, IL 60637, USA*

## Abstract

The Extreme Universe Space Observatory on a Super Pressure Balloon 2 (EUSO-SPB2), in preparation, aims to make the first observations of Ultra-High Energy Cosmic Rays (UHECRs) from near space using optical techniques. EUSO-SPB2 will prototype instrumentation for future satellite-based missions, including the Probe of Extreme Multi-Messenger Astrophysics (POEMMA) and K-EUSO. The payload will consist of two telescopes. The first is a Cherenkov telescope (CT) being developed to quantify the background for future below-the-limb very high energy (E>10 PeV) astrophysical neutrino observations, and the second is a fluorescence telescope (FT) being developed for detection of UHECRs. The FT will consist of a Schmidt telescope, and a 6192 pixel ultraviolet camera with an integration time of 1.05 $\mu$s. The first step in the data acquisition process for the FT is a hardware level trigger in order to decide which data to record. In order to maximize the number of UHECR induced extensive air showers (EASs) which can be detected, a novel trigger algorithm has been developed based on the intricacies and limitations of the detector. The expected performance of the trigger has been characterized by simulations and, pending hardware verification, shows that EUSO-SPB2 is well positioned to attempt the first near-space observation of UHECRs via optical techniques.

## 1. Introduction

Ultra high energy cosmic rays (UHECRs) provide a window into the most energetic phenomena of the observable Universe. By studying these particles, we can hope to better understand their creation mechanisms and gain insight into the most extreme astrophysical environments (Sarazin et al., 2019; Murase et al., 2012). With energies more than an order of magnitude greater than what can be recreated in laboratory settings on Earth, they also provide an avenue to study particle interaction models at the highest energies (Valiñ, 2015). The flux of cosmic rays falls rapidly with energy with $S(E) \propto E^{-2.5}$ in the range 5 to 13 EeV (Aab et al., 2020). The rapidly falling energy spectrum combined with the flux scale, make it not possible to study cosmic rays through direct measurement when their ener-gies exceed $10^{15}$ eV. To overcome that, indirect techniques have been developed. When an UHECR interacts with the atmosphere, a cascade of secondary particles is created called an extensive air shower (EAS). The majority of energy deposited by the primary particle will ultimately become electromagnetic cascades which produce ultraviolet fluorescence light from collisonally excited nitrogen in the air. By observing these photons characteristics of the primary particle can be studied.

* Corresponding authors at: Universitá degli Studi di Torino via P. Giuria 1, 10125 Torino, Italy (M. Battisti).
*E-mail addresses:* gfilippatos@mines.edu (G. Filippatos), matteo.battisti@edu.unito.it (M. Battisti).

Since UHECRs at the highest energies are incredibly rare, with one particle above $10^{19}$ eV arriving roughly once per km$^2$ per year, large detectors have been built in order to study them. The Pierre Auger Observatory (Aab et al., 2015) covers 3000 km$^2$ in the southern hemisphere and the Telescope Array (TA) (Abu-Zayyad et al., 2012) covers 700 km$^2$ in the northern hemisphere. These two experiments have greatly advanced our understanding of UHECRs over the two decades since they have been deployed. Nonetheless, open questions still remain. These include the nature of the flux suppression observed at the highest energies, the mass composition of UHECRs and different locations of hot spots and anisotropy observed by Auger and TA (Aab et al., 2017; Abbasi et al., 2020). In order to pursue answers to these questions, a UHECR detector with a large observation aperture and uniform sky coverage is needed. One proposed way to achieve this is by utilizing a space-based detector, proposed by J. Linsley in 1980s and developed in the framework of the JEM-EUSO collaboration (Benson and Linsley, 1981; Casolino et al., 2017) with a first attempt to measure UHECR from space by the TUS experiment (Khrenov et al., 2020; Klimov et al., 2017).

The aim of the JEM-EUSO program (Ricci, 2016) is the study of UHECRs through a space-based instrument, with a large field of view and uniform coverage over both hemispheres. Its observation principle is based on the detection and measurement of the fluorescence light produced by the EASs developing in the atmosphere. Presently in preparation is the second iteration of EUSO on a long duration balloon flight, EUSO-SPB2. With the goal of building on previous EUSO missions and preparing for future satellite based missions, EUSO-SPB2 aims to make several landmark observations. The payload features two telescopes, a fluorescence telescope (FT) and a Cherenkov Telescope (CT). EUSO-SPB2 will prototype both portions of the focal surface for the planned Probe of Extreme Multi-Messenger Astrophysics (POEMMA) (Olinto et al., 2021). EUSO-SPB2 is currently scheduled for launch in 2023 from Wanaka, New Zealand on board a NASA super pressure balloon (Zell, 2017).

The FT will attempt observations of EASs via fluorescence similar to previous JEM-EUSO balloon missions (Wiencke and Olinto, 2017; Abdellaoui et al., 2018). With more advanced optics and electronics, as well as a more mature data acquisition (DAQ) system than previous missions, the FT is designed to make the first optical observation of EAS tracks from above.

The FT will feature two parallel data pipelines, a single photon counting mode and a continuous integration mode. In standard photon counting mode, two photo-electrons (PEs) detected in the same pixel within the time window of the double pulse resolution (∼6 ns) will be counted as one, as the two PE peaks will be indistinguishable from one another. In the most extreme scenario, a very bright source with PEs being generated continuously with less than 6 ns between each, no peaks would be identifiable and would result in zero counts being recorded for that pixel during that integration. For observing typical EAS, this is generally not a major issue since the energy required to produce a bright enough signal is larger than the highest energy UHECRs ever observed.

The Cherenkov light produced by a cascade of particles moving towards the optical system of the detector will however reach such extreme intensities. This kind of events might be generated by upgoing showers initiated by very high energy (E>10 PeV) tau neutrinos.

Given the observational geometry of the fluorescence telescope, pointing downwards, signals from these types of events are not expected to be observed. However, anomalous events have been observed by the ANITA experiment, an Antarctic balloon-borne detector observing EAS through radio frequencies (Gorham et al., 2016; Hoover et al., 2010). The explanations for these events range from beyond standard model particles (Fox et al., 2018) to dark matter annihilation events (Liang and Zhitnitsky, 2021). The fluorescence telescope on-board EUSO-SPB2 may be capable of providing valuable insights into the feasibility of these scenarios. Due to the expected signal being brighter than what the single photon counting is sensitive to, the continuous integration mode will be necessary for measurements or non-measurements of this type of signal to be verified.

In Section 2 we will briefly discuss the detector at the heart of the FT, in Section 3 the general structure of the data acquisition system is presented, in Section 4 the details of the trigger will be summarized, Section 5 will present the methodology and results of the simulations, while Section 6 will discuss the tests planned and performed. Finally, Section 7 will present the concluding remarks.

## 2. Detector description

The EUSO-SPB2 FT builds on the experience of previous missions, but will have several distinct advantages. One major improvement will be the optics used. Rather than Fresnel lenses utilized on previous EUSO balloon flights (Scotti and Osteria, 2016; Osteria et al., 2019), EUSO-SPB2 will fly Schmidt telescopes with entrance apertures of 1 m. Consisting of six spherical mirror segments with a radius of curvature of 1.6 m and an aspheric corrector plate to account for spherical aberrations, the Schmidt optics are designed to improve both the collecting power and focusing capability of the instrument. Overall, the optical throughput of the instrument is expected to be 68% and the point spread function (PSF) is expected to contain 95% of the light within 3 mm. The FT is visible inside the EUSO-SPB2 gondola in Fig. 1.

The focal surface of the FT will consist of three photo-detection modules (PDMs) one of which is shown in Fig. 2. This will be the first EUSO mission to consist of multiple PDMs. These PDMs each consist of nine elementary cells (ECs), each of which is made up of four multi-anode photo-multiplier tubes (MAPMTs Hamamatsu Photonics R11265-M64), for a total of 2304 pixels arranged in a square matrix. Each of the nine ECs shares a common high voltage power supply (HVPS) based on a Cockroft-Walton circuit. The ECs are compact assemblies each containing one HVPS generator board and an application-specific integrated circuits (ASICs) for signal digitisation. This assembly is potted in a gelatinous compound to prevent discharge between the various components. The EC assemblies are roughly 55 mm×55 mm. The MAPMTs operate in photon counting mode, with an integration time of 1.05 $\mu$s which is defined as 1 Gate Time Unit (GTU). Each MAPMT is made up of 64 $3.88 \times 3.88$ mm$^2$ pixels arranged in an 8×8 grid. The quantum efficiency of the MAPMTs is roughly 33% in the wavelength range of interest, with a collection efficiency of 80%. The total field of view (FoV) of the instrument is around $12° \times 36°$. Projecting this onto the ground at a float altitude of 33 km results in an area of 36 km$^2$. Each individual pixel has a FoV of ~0.25°. There are gaps between PDMs, ECs and MAPMTs which result in parts of the atmosphere which cannot be observed.

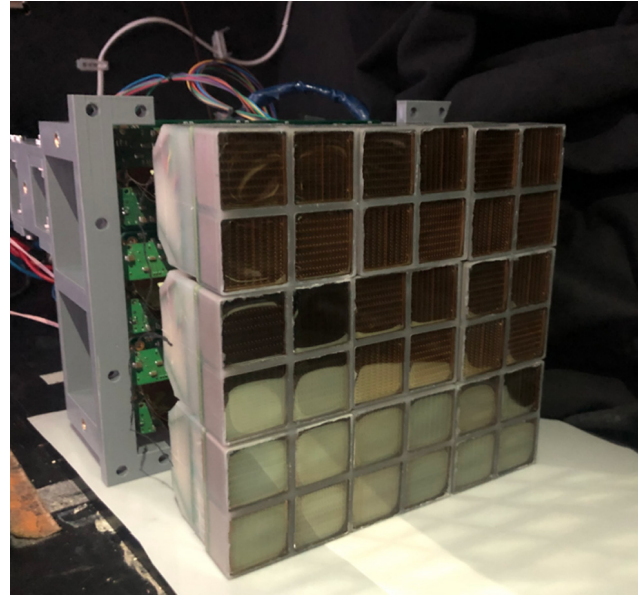

Fig. 2. Photograph of an assembled PDM, of which the FT will consist of three side by side. Four MAPMTs are potted into gelatinous gel to form an EC. These ECs are then connected to a cross-board, the green printed circuit boards in the back. The whole PDM is enclosed in a custom plastic enclosure.

## 3. Data Acquisition System

The EUSO-SPB2 data acquisition system is derived from the Mini-EUSO (Belov et al., 2018) and EUSO-SPB1 (Scotti et al., 2019) systems and works on a PDM-basis. Each MAPMT is read out by a SPACIROC3 ASIC (Blin et al., 2018) that performs both photon counting and continuous integration. The photon counting is the main mode of the FT. For each of the 64 pixels in a MAPMT the data is digitized in acquisition windows of 1 GTU, with a double pulse resolution of ~ 6 ns. The continuous integration is performed integrating the signal from 8 pixels. The data is digitized and later used for the research of instantaneous bright events.

The output of the 36 ASICs of a PDM is collected by three cross boards, each of which contains one Artix 7 FPGA. The cross boards perform data gathering from the ASICs and data multiplexing. The three cross boards are connected to the main electronic board, referred to as a Zynq board, containing a Xilinx Zynq 7000 FPGA with an embedded dual core ARM9 CPU processing system. There is one Zynq board for each PDM. The Zynq board controls the data flow from the ASICs, runs the trigger logics (for the photon counting and charge integrated data), and interfaces with the external CPU for the data storage.

When a trigger is issued, a signal is sent to an external board, called the Clock board, that accepts signals from the three Zynq boards and generates its own trigger signal sent back in parallel to the Zynq boards of the other PDMs. The trigger may come either from the photon counting, or from the continuous integration data. The Clock board also maintains the time synchronization of the entire telescope and performs the timing and position tagging of each triggered event using data from two GPS receivers. Upon a trigger signal, 128 GTUs are stored, 64 GTUs before and 64 GTUs after the trigger, from both the photon counting and the charge integrated data. Together with the data, other ancillary information are stored.

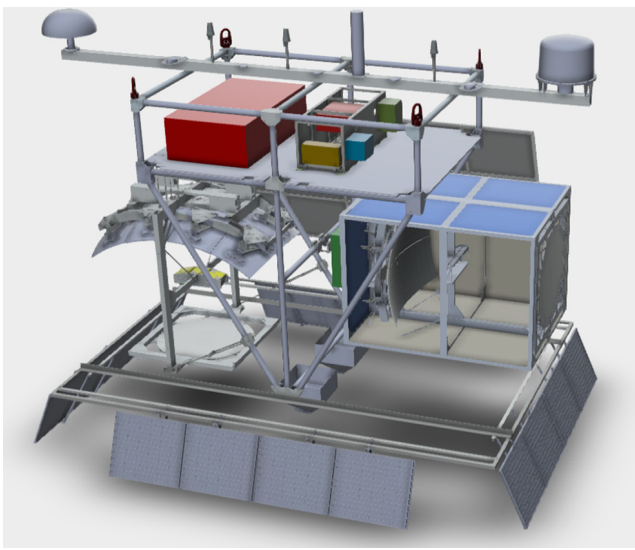

Fig. 1. CAD rendering of the EUSO-SPB2 gondola including the two telescopes. The left shows the FT pointing downward. The right shows the CT, with the light-tight clam shell, which will surround both telescopes (open side for visualization purposes only). Solar panels are shown around the bottom and on the back. Control computers and telemetry equipment are on the top of the structure.

The DAQ process for an individual PDM is summarized in Fig. 3.

## 4. L1 Trigger Logic

### 4.1. Trigger Logic Requirements

Once in flight, the FT is expected to detect several types of signals above threshold. Most of them will be background, therefore an online trigger is designed to recognize and flag every event that might be produced by an EAS developing in the atmosphere. This trigger should be able to catch all obvious events, such as the example depicted in Section 5, while also catching as many other EASs as possible. A particular focus has been to reduce the trigger energy threshold, since the ability to measure lower energy cosmic rays will lead to significantly more observable events, given the nature of the cosmic ray energy spectrum. Finally, given the bandwidth and computational resources available for the EUSO-SPB2 mission, the trigger logic is required to produce a trigger rate of $\lesssim 1$ Hz/PDM.

To achieve these results, the trigger logic has been designed taking into account the specificities of the detector (pixel size, altitude of operation, GTU length) as well as the expected characteristics of the EAS signals. The speed at which an EAS moves across the camera is related to the properties of the detector and the geometry of the shower. An EAS pattern can be described by two angles: zenith ($\theta$) which is the angle with respect to a vertical line perpendicular to the Earth's surface, and azimuth ($\phi$) which is related to the choice of coordinate system. For a nadir-looking detector like EUSO-SPB2, the direction and speed with which the shower crosses the camera, are related to $\phi$ and $\theta$ respectively. For a shower crossing directly below the center of the detector, the angular speed with which it will cross the camera is given by

$$u = \frac{c \sin \theta}{\alpha D} \tag{1}$$

where $\alpha$ is the observation angle of a pixel, $D$ is the distance from the shower to the detector and $c$ is the speed of light.

As can be seen by Fig. 4, the majority of showers crosses more than one pixel during each GTU, spreading the signal over several pixels and therefore decreasing the signal-to-noise ratio. For a space based mission, a 2.5 $\mu$s GTU is appropriate, for an SPB flight lower would be ideal due to the lower orbital altitude. Using the electronics designed for future space based missions, 1.05 $\mu$s is as low as can be achieved. To overcome this problem, the trigger logic is designed to work at "macro-pixel" level, a macro-pixel being a 2x2 group of non overlapping pixels, transforming a PDM from a 48×48 pixel detector into a 24×24 macro-pixel detector.

The benefit of this approach is illustrated in Fig. 4 where it can be seen that the signals from showers with a zenith angle up to $\theta = 50°$ do not cross more than one macro-pixel in a single GTU, assuming the shower develops at 10 km altitude. This results in the signal to noise ratio being larger within macro-pixels than individual pixels for these types of showers. For a more general set of showers, the right panel of Fig. 4 shows some showers that can be expected to cross up to four pixels in a single GTU.

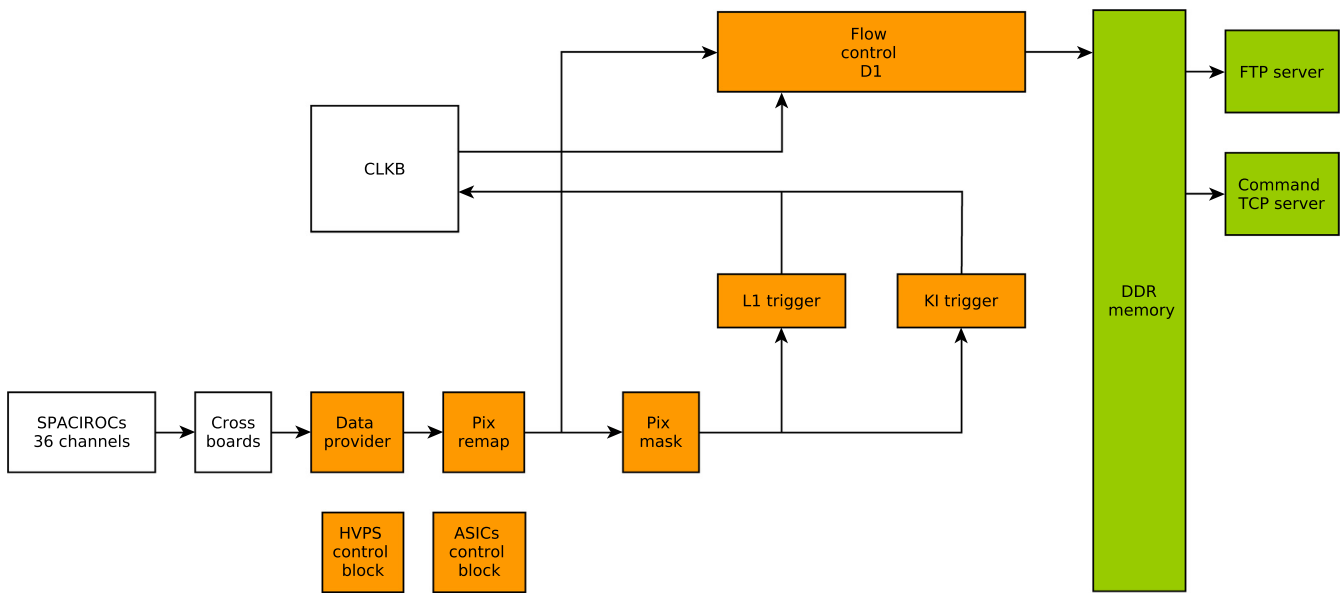

Fig. 3. EUSO-SPB2 data flow for a single PDM. Data coming from the 36 ASICs (SPACIROCs) is multiplexed by the cross boards and sent to the Zynq board FPGA, the orange blocks in this diagram. Potential malfunctioning or noisy pixels can be prevented from triggering (pixel masking), independently for the photon counting and charge integrated data. The data is sent to the L1 trigger (photon counting) and to the KI trigger (continuous integration). When given conditions are satisfied, a trigger is issued. The trigger signal is sent to the Clock board (CLKB) that sends back an external trigger to the Zynq boards controlling the other PDMs. When the trigger is received, 128 GTUs from all the 3 PDMs are stored in the DDR memory. This includes the 64 GTUs before and 64 GTUs after the trigger signal. In addition, some ancillary data is stored simultaneously. Data is transferred to the FTP server which outputs data as files. The Zynq board also controls the high voltage power supply (HVPS) and configures the ASICs.

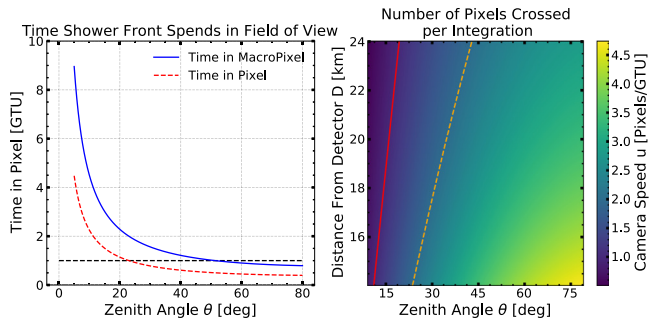

Fig. 4. Time that a shower at 10 km altitude spends in one pixel and one macro pixel as a function of zenith angle (left). Angular speed of shower as seen in the detector as a function of distance of the shower from the detector and the zenith angle of the shower, as described in Eq. 1 (right). The red line shows where the camera speed is equal to 1 pixel/GTU. The dashed gold line shows where the camera speed is equal to 2 pixel/GTU, or 1 .macro-pixel/GTU.

### 4.2. Trigger Logic Design

As already mentioned, the trigger logic works at macro-pixel level. In general, it looks for an area of the PDM where a cluster in space and time of macro-pixels over threshold is present.

Each macro-pixel has its own independent threshold based on the average count rate observed by that macro-pixel. The threshold per macro-pixel $i$ is updated twice per second, and its value is given by

$$\text{THR}_i = n_\sigma \sqrt{\lambda_i} + \lambda_i \quad (2)$$

where $\lambda_i$ is the average count rate in the macro-pixel $i$ over the 16384 GTUs prior to the threshold update, and $n_\sigma$ is a parameter that controls the final threshold value. Assuming the counts follow a Poisson distribution, the standard deviation is equal to the square root of the average, therefore the threshold is set $n_\sigma$ standard deviations above the average. The minimum value for the threshold is set to correspond to the threshold of a macro-pixel with $\lambda_i = 0.5$ counts/GTU. This value can be changed during flight to keep the trigger rate at an acceptable level.

The values of $\lambda_i$ are stored and downlinked to ground, and will be used to monitor the thresholds, to produce a long-exposure datastream at a tens of ms timescale and, most importantly, to compute the final exposure of the instrument. The choice to update the thresholds every 500 ms is motivated by the fact that a stratospheric balloon moves at $\sim 100$ km/h, i.e. $\sim 15$ m every 500 ms. Therefore, the field of view of a macro-pixel changes only $\sim 5\%$ in 0.5 s. Moreover, EUSO-SPB2 will fly over the Southern Ocean and is not expected to spend significant time over continents where bright light sources like cities are present. These considerations guarantee that the threshold's update is fast enough to prevent triggers from non-flashing, diffused sources, for example reflecting high clouds or the rising moon.

In parallel to the threshold update process, a binary matrix of macro-pixels over threshold is calculated every GTU. Every element of the $24 \times 24$ binary matrix indicates whether a macro-pixel is above threshold. The remaining part of the trigger logic takes only the binary matrix into account, not the values of single macro-pixels. This has the advantage of preventing one overly bright flash, which is not indicative of an EAS, from triggering. Using the binary matrix, clusters of macro-pixels over thresholds in space and time are searched for. If the number of macro-pixels within a $3 \times 3$ grid, and 3 GTUs of time is more than $n_{\text{hot}}$ then that cluster is considered active. Macro-pixels on the border of the PDM are not considered, as they do not have a full grid surrounding them. A trigger is issued when more than $n_{\text{active}}$ clusters are active within $l$ GTUs. A step-by-step description of the trigger logic is shown in Fig. 5.

The trigger logic has therefore 4 parameters: $n_\sigma, n_{\text{hot}}, n_{\text{active}}, l$.

This approach to the trigger logic presents definite advantages with respect to the logic adopted for the previous balloon flight, EUSO-SPB1, and adapted from the first level trigger of the JEM-EUSO experiment (Abdellaoui et al., 2017). Despite its good performance both in simulation and during field tests (Battisti et al., 2019), the previous trigger logic had the major drawback to set the same threshold for all the 64 pixels of a MAPMT, often resulting in an artificial and uncontrolled increment of the threshold for some pixels.

In addition to the photon counting trigger described above, there is a continuous integration trigger designed to detect anomalously bright signals. The continuous integration works by grouping clusters of eight pixels per MAPMT together and integrating the charge detected on them continuously rather than searching for discrete pulses as is the case with the photon counting mode. This allows for much brighter signals to be observed such as those generated from Cherenkov cones around upward going air showers.

The continuous integration channel will be read out in parallel to the photon counting channel each time a trigger is issued. Additionally, there is dedicated continuous integration trigger. This is much simpler than the photon counting trigger, and is issued when a continuous integration channel is greater than a given threshold. When the continuous integration trigger is issued, the single photon data is recorded as well.

### 5. Simulations

The trigger algorithm has been tested using extensive simulations. These start with an EAS generated by simulat-ing the interactions of the primary particle and continuing the cascade down to all particles of relevant energy. The simulations of the EAS are done using the program Conex (Bergmann et al., 2007). Once the EAS is generated, the UV light produced is then propagated through the atmosphere in order to determine how much light will arrive at the detector. After this the response of the detector is simulated by using the Geant4 framework (Agostinelli et al., 2003) in order to mimic the response of the instrument. The propagation of the shower through the atmosphere as well as the response of the detector are carried out using the JEM-EUSO Offline framework (Paul, 2015).

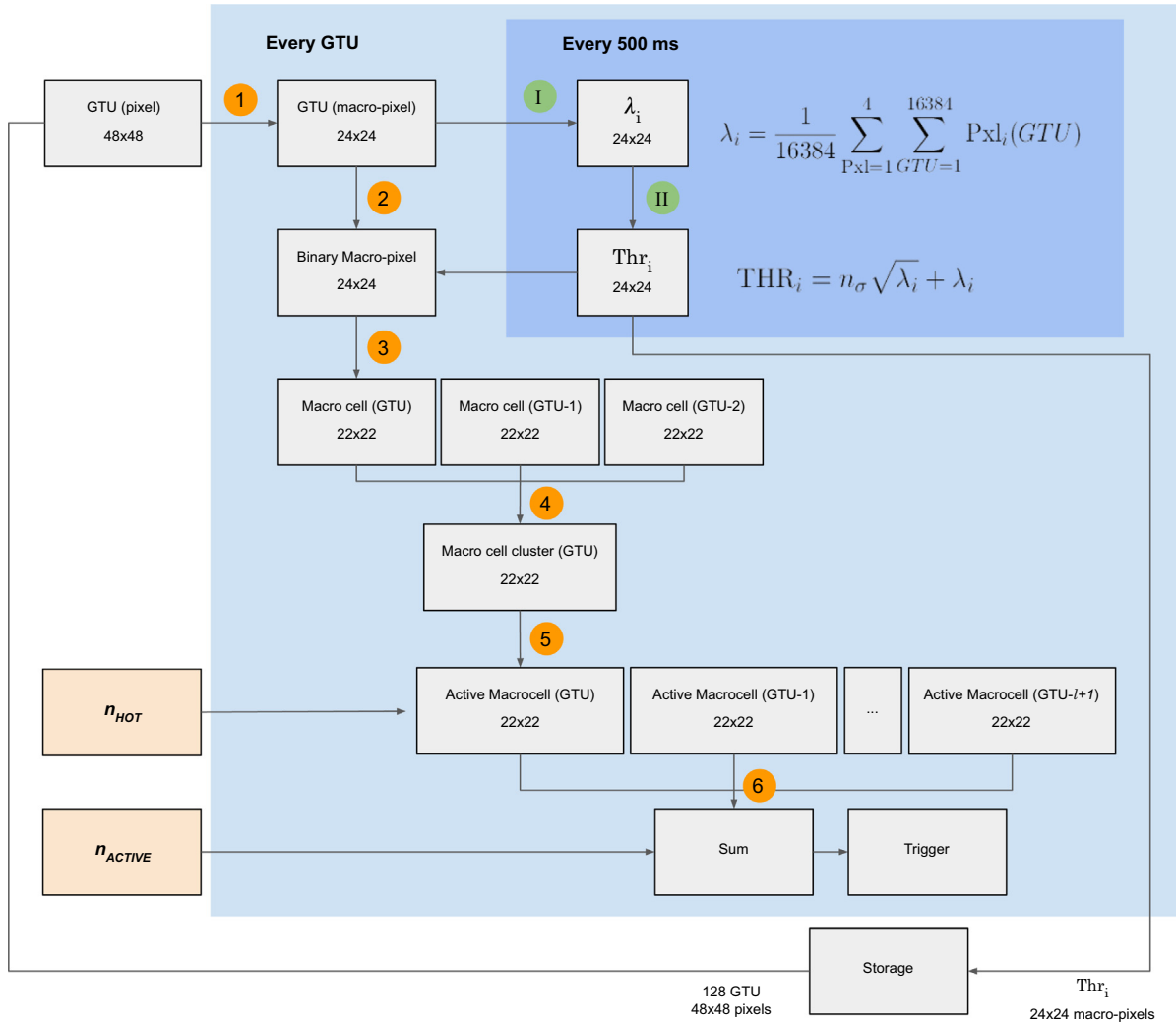

Fig. 5. Trigger scheme of FT trigger logic, every GTU: 1) The values of the macro-pixels are computed, obtaining a 24 × 24 view of the PDM. 2) The Binary Matrix is created through a comparison between each macro-pixel and its threshold. 3) The PDM is divided into 22 × 22 overlapping 3 × 3 macro-cells, excluding macro-pixels on the border of the PDM, the value of each cell is the number of macro-pixels over threshold. This matrix is stored in a 3 slots FIFO circular buffer, containing the values for the current and the two previous GTUs. 4) The sum over the 3 GTUs is performed. Each element of the Cluster matrix contains the number of macro-pixels over threshold in the last 3 GTUs in each 3 × 3 macro-cell. 5) The number of clusters with more than $n_{hot}$ macro-pixels over threshold is stored in a $l$ length FIFO circular buffer. 6) The total number of active macro-cells over the last $l$ GTUs is computed and compared to the value of $n_{active}$. If SUM > $n_{active}$ a trigger is issued. In parallel, every 500 ms I) the average of each macro-pixel over the previous 16384 GTUs is computed. II) This value is used to compute the threshold for each macro-pixel.

An example of a simulated shower is shown in Fig. 6. The simulated photons in the field of view (FoV) are shown in red, and the PEs detected are shown in blue. As visible in this image from the fraction of PEs compared to total photons in the FoV, an overall efficiency of 18% is reproduced, which is the current expected end-to-end throughput of the FT. The gaps in the blue histogram show the effect of the gaps in between MAPMTs and ECs. The focal plane view of the same event is shown in Fig. 7 where the track produced by the EAS is clearly visible. In order to interpret how various aspects of the FT affect its ability to detect EAS we often look at the impact on the expected event rate. The number of expected EAS observations is dependent on a variety of factors and is estimated by using a large Monte Carlo simulation. Showers are thrown uniformly over a large area below the detector. Angles and core locations are chosen to mimic an isotropic flux of incoming showers, with $f(\theta) \propto \cos\theta \sin\theta$ and $f(r) \propto r^2$ over a 100 km radius disk and zenith angles from 0° to 80°. Primary particles are protons generated in twenty energy bins evenly spaced in $\log_{10}(E/eV)$ from $E = 10^{17.8}$ eV to $E = 10^{19.7}$ eV. Ten thousand showers per energy bin are thrown eight times each for a total of 1.6 million simulated EAS. The fraction of triggered events for a given energy $f(E_i)$ is then converted into an event rate by using the energy spectrum as measured by the Pierre Auger Observatory (Aab et al., 2020) via Section 5,

$$R(E_i) = f(E_i) A \Omega \int_{E_i - \Delta E/2}^{E_i + \Delta E/2} J(E) dE. \quad (3)$$

where A is the area over which showers are thrown and Ω is the solid angle simulated given by $\Omega = \int_0^{80°} \sin\theta \cos\theta \, d\theta$ and $J(E)$ is the differential energy spectrum. The total event rate is then given by the sum $\sum_{i=1}^{20} R(E_i)$. An example of this is shown in Fig. 8. The relatively small fraction of triggered events is a result of an oversized area being sampled, this number should not be thought of as a trigger efficiency.

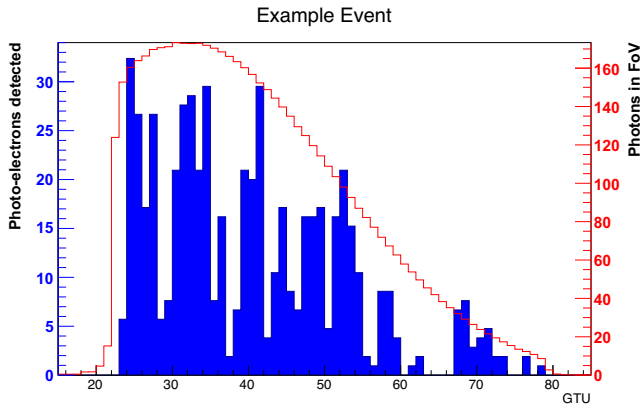

Fig. 6. Simulated shower with Energy E = 3 EeV, zenith angle θ = 57°, and ideally placed below detector. Blue histogram shows the number of recorded photo-electrons per GTU, across all pixels generated by light from the EAS. Red histogram shows the photons which arrive at the aperture of the detector per GTU.

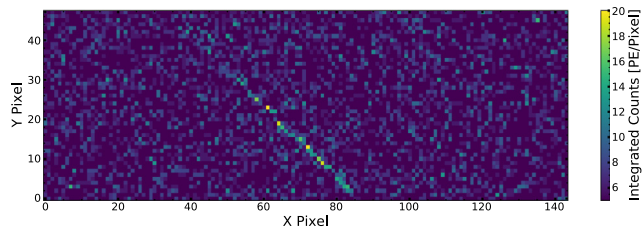

Fig. 7. Photo-electrons (PEs) per pixelm integrated over 50 GTUs, shower with Energy E = 3 EeV, zenith angle θ = 57°, in an ideal location for observation below detector as in Fig. 6. Simulated background of 0.1 average counts per pixel per GTU, which is lower than the expected background rate in flight but is useful for illust.rative purposes.

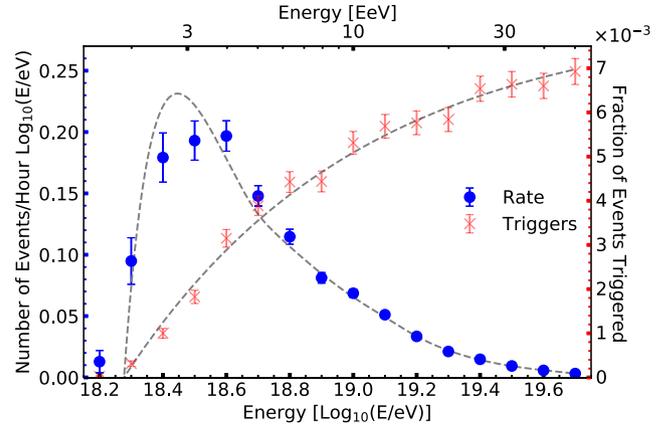

Fig. 8. The fraction of triggered showers per energy bin shown in red, fit to an activation function of the form $a(1 - \exp(-(x+b)/c))$. The estimated event rate as a function of shower energy shown in blue, normalized to be differential with respect to $\log_{10}(E/eV)$. Error bars shown are Poissonian errors on the number of triggered events. Image taken from (Filippatos et al., 2021).

Via simulation, the trigger algorithm can be simulated quickly allowing for the wide range of parameters to be investigated. The four control parameters provide a very wide range of control over the trigger in unique ways.

There is a physical motivation for the parameter $n_\sigma$ to be as low as possible. Due to the nature of the UHECR energy spectrum, detecting events with a sightly lower energy can lead to a large increase in the total number of observed events. With this in mind, a range of values for each parameter were investigated:

- $n_\sigma \in [2, 2.5, 3, 3.5, 4, 4.5, 5, 5.5, 6, 6.5]$
- $n_{hot} \in [2, 3, 4, 5, 6, 7, 8, 9]$
- $n_{active} \in [3, 5, 8, 13, 21, 34]$
- $l \in [5, 10, 15, 20]$

In total 1920 trigger combinations were tested on both simulated background and the simulated EASs. The background used followed a Poisson distribution for each pixel, and represented ten seconds of real data taking. Combinations that found no false triggers in this simulated ten seconds of data were considered strict enough. This selects combinations that yield a false trigger rate of ≲ 0.1 Hz/ PDM. This is more strict than required, however we expect many triggers in addition to those due to poissonian fluctuations, such as direct hits from low energy cosmic rays. Of these, the combination that led to the largest number of triggered showers was $n_\sigma$=5, $n_{hot}$=2, $n_{active}$=34, $l = 20$. The effects of varying the four control parameters around the chosen values are shown in Fig. 9. Here, two trigger parameters are varied at a time, and the other two are kept at the ideal values mentioned above. The color shows the event rate relative to the event rate given the ideal parameters. The white hashed regions show the parameter combinations which result in a false trigger rate that is higher than allowed by the on-board electronics and available bandwidth.

As can be seen, the impact on the event rate is not uniform across the parameters. For example, the impact of the $n_{hot}$ parameter is much more significant than the impact of the $l$ parameter. Further, several combinations yield nearly identical rates of EAS observation. By dynamically changing trigger parameters to respond to varying observing conditions, we can attempt to minimize the incidence of background triggers while maximizing the number of observed EAS.

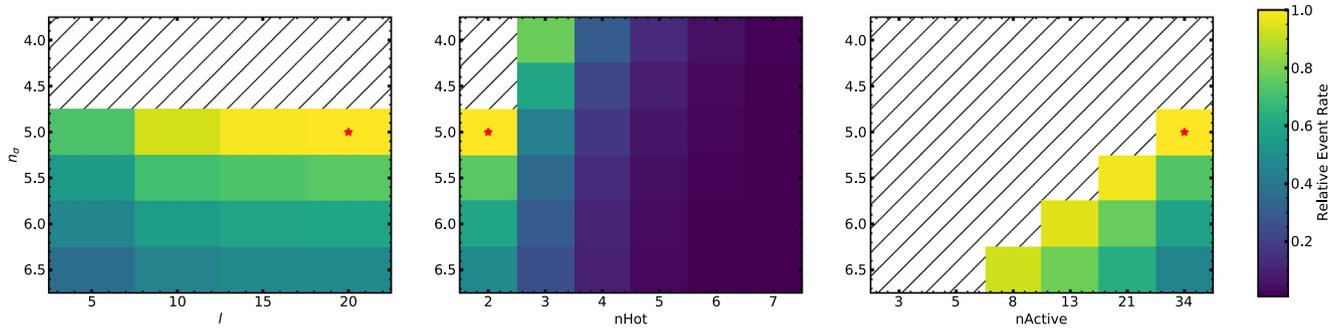

Fig. 9. Impact on expected event rate of varying the trigger parameters. Two parameters varied per plot, $n_\sigma$ and one of the others. Common color scale shown to the right represents the relative event rate. This is found by normalizing the expected rate to the expected rate given $n_\sigma = 5, n_{\text{hot}} = 2, n_{\text{active}} = 34, l = 20$. White hashed regions of the above plots show configurations of parameters that are disallowed due to the trigger rate on poissonian background being too high at $> 0.1$ Hz/PDM. Locations of ideal trigger parameters shown by the red stars.

## 6. Pre-Flight Validations

### 6.1. Laboratory Tests

The proposed trigger logic and the whole data acquisition system will be thoroughly tested before launch. Functional tests will be carried out in different environmental conditions to control the correct implementation and functioning of the data acquisition and trigger system, and to test the ability of the trigger logic to correctly detect light signal at the μs timescale. The tests will be performed first on a scaled-down version of the system, consisting in only one EC (16×16 pixels), with the exact same electronic of the final configuration but without the optical system. These tests will be mainly carried out at the TurLab Facility, a laboratory hosted in the Physics Department of the University of Turin. The TurLab facility is a laboratory equipped with a 5 m diameter and 1 m depth rotating tank, located in an underground level of the Physics Department of the University of Turin. The lab was originally built to perform geo-fluid-dynamics experiments, taking advantage of the rotating tank that can be operated at a speed ranging from 3 s to 20 min per rotation. It also gives the possibility to artificially control the light intensity. In the past years, the lab has been used to test the data acquisition system and the trigger logic of previous JEM-EUSO missions, namely EUSO-SPB1 (Suino et al., 2017) and Mini-EUSO (Bisconti et al., 2021). The detector is hanged from the ceiling above the rotating tank, in a structure that allows the use of a small plano-convex lens as the optical system, the light level is set to the desired level (usually a level that gives the expected background from moonless night, ∼ 1 count/pixel/GTU) and different light sources are placed inside the tank. Rotating the tank, it is possible to mimic the movement of a balloon-based or space-based detector as it observes the Earth atmosphere while different sources enter its field of view. The sources can either be passive, i.e. different materials that reflect the light in different way, or active, like for example flashing LEDs mimicking lightning strikes or an Arduino-driven strip of LEDs mimicking EASs. The TurLab facility is also a perfect location for static measurement, as it is housed in a very large chamber more than 40 m long, orders of magnitude darker than the night sky.

These tests will check the implementation of the data acquisition system in different environmental conditions, as well as the correct implementation and operation of the trigger logic.

### 6.2. Field Tests

Field tests are scheduled for 2022. Similar to the EUSO-SPB1 field tests (Adams et al., 2021), the EUSO-SPB2 field tests will utilize a laser of known energy to preform an end to end calibration of the instrument (Hunt et al., 2016). These field tests will also crucially serve as a metric to benchmark the performance of both the trigger and the JEM-EUSO Offline framework, which has the ability also to simulate lasers. These tests will allow for necessary refinements to be made to the instrument.

## 7. Conclusions

Serving as a pathfinder instrument, EUSO-SPB2 provides an opportunity to learn about the challenges necessary for observing UHECRs from near-space. The data acquisition pipeline is complex and requires many pieces to work together and to do so at microsecond timescales. While it is only one part of this pipeline, developing an efficient hardware level trigger which can intelligently distinguish between EAS induced signals and other sources of light is essential to the success of EUSO-SPB2. The same will be true for future orbital and sub-orbital missions. By tailoring a trigger algorithm to the specific design of the instrument, we hope to maximize the scientific impact of the experiment by maximizing the number of UHECR-induced EAS which can be observed.

Utilizing extensive end to end simulations from the primary interaction to the detector response, the trigger algorithm can be both tested and optimized. Prior to launch, these simulations along with the validity of the trigger methodology will be able to be tested via laboratory measurements and field tests. Aided by this trigger, EUSO-SPB2 is well positioned to provide valuable insights for future space-based experiments, and attempt the first near-space observation of an EAS via nitrogen fluorescence.

**Declaration of Competing Interest**

The authors declare that they have no known competing financial interests or personal relationships that could have appeared to influence the work reported in this paper.

**Acknowledgments**

This work was partially supported by NASA Grant No.: 80NSSC18K0477, the Interdisciplinary Scientific and Educational School of Moscow University "Fundamental and Applied Space Research", by the Italian Ministry of Foreign Affairs and International Cooperation, by the Italian Space Agency through the ASI INFN agreements n. 2017–8-H.0, n. 2021–8-HH.0. This research used resources of the National Energy Research Scientific Computing Center (NERSC), a U.S. Department of Energy Office of Science User Facility operated under Contract No. DE-AC02-05CH11231. The authors would like to thank the JEM-EUSO members for their inputs, and in particular the EUSO-SPB2 group.